# ALS STORAGE RING RF CONTROL SYSTEM UPGRADE PLAN AND STATUS*

N. Us Saqib†, A. Jurado, E. Andrade, Q. Du, J. H. Lee, M. Dach, B. Flugstad
Lawrence Berkeley National Laboratory, Berkeley, CA, United States

*Abstract*

The Advanced Light Source (ALS) at Lawrence Berkeley National Laboratory, a third-generation synchrotron light source operational since 1992, is undergoing a comprehensive upgrade of its storage ring RF control system. The legacy Horner PLC controllers and remote I/O modules, now at end-of-life, are being replaced with an Allen-Bradley PLC platform to improve maintainability, reliability, and long-term support. This paper presents the planning, design, and current status of the upgrade project.

## INTRODUCTION

The Storage Ring (SR) Radio-Frequency (RF) system at the Advanced Light Source (ALS) is a critical subsystem that delivers the RF power required to accelerate and sustain the 1.9 GeV electron beam. It consists of two single-cell RF cavities operating at 499.642 MHz, driven by a pair of klystrons [1]. While the core storage ring RF hardware—including cavities, waveguides, circulators, and high-power switches—remains central to the ongoing ALS-U upgrade project, the associated end-of-life Horner PLC-based control system represents an operational risk, creating substantial challenges for long-term reliability and maintainability of the SR RF control system.

To address this, a phased upgrade plan has been developed to replace the existing PLC controls with a modern, maintainable architecture based on Allen-Bradley PLCs, consistent with ALS-U PLC-based systems [2]. The upgrade is executed in multiple phases, each designed for completion within a two-month shutdown to minimize operational disruption. This paper presents the current SR RF control system, the upgrade plan and status, and the design of the new system.

## CURRENT SR RF CONTROL SYSTEM

The SR RF control system manages the monitoring and control of the storage ring RF system according to its operational configuration. The SR RF system can be configured in six modes which are essential for the control system to manage operation and enforce interlocks effectively:

1. Dual Klystrons: Klystron 1 drives Cavity 1 and Klystron 2 drives Cavity 2.
2. Klystron 1 Drive: Klystron 1 to both cavities, power split via the magic Tee.
3. Klystron 2 Drive: Klystron 2 to both cavities, power split via the magic Tee.
4. Klystron 1 Test: Klystron 1 to dummy loads.
5. Klystron 2 Test: Klystron 2 to dummy loads.
6. Cavity Measurement: Low-power VNA connections to the cavities.

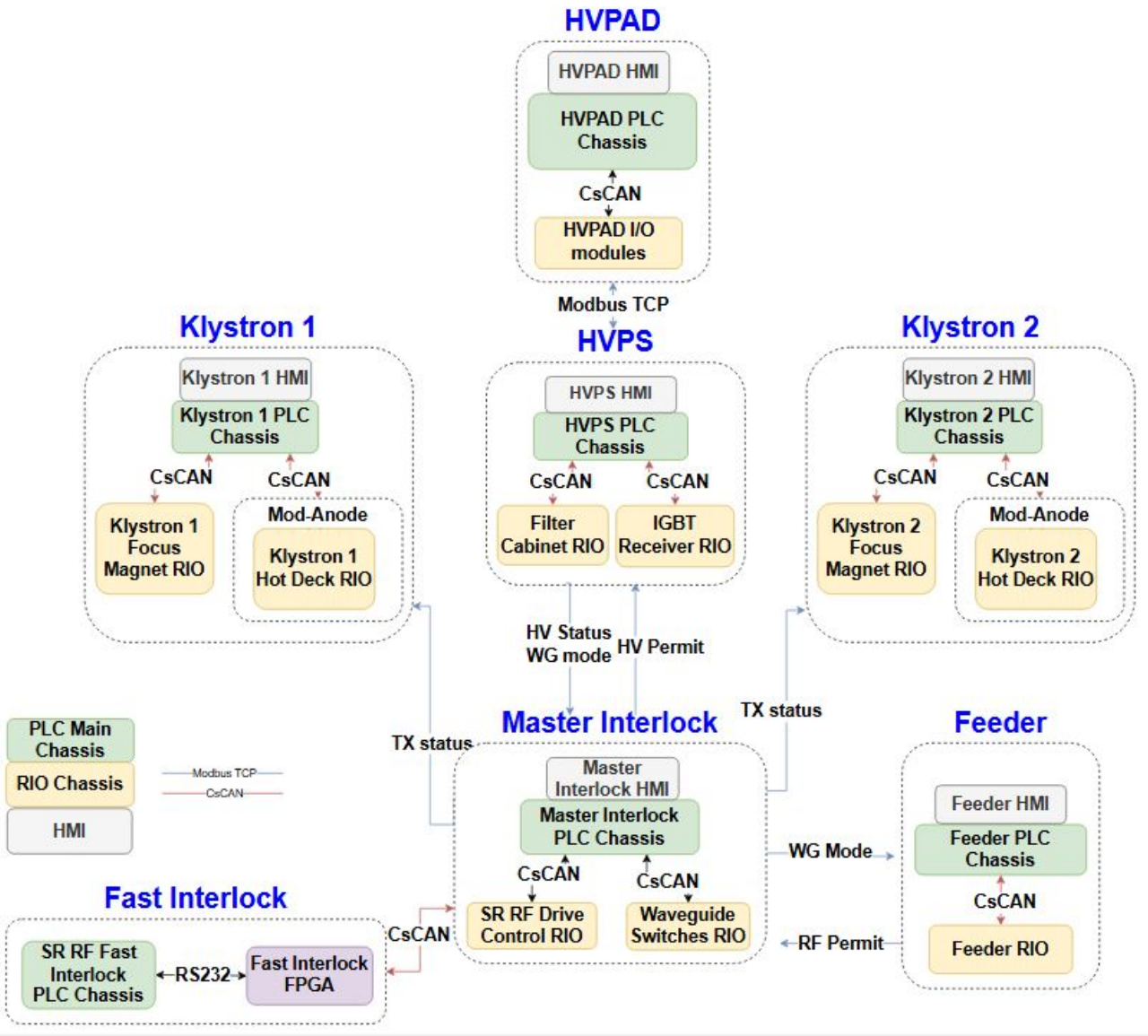

Figure 1: Existing system architecture.

Figure 1 shows the architecture of the existing SR RF PLC-based control system, which comprises the following subsystems:

- High Voltage Power Supply (HVPS) subsystem – Selects configuration mode, controls high-voltage regulation and manages sequencer-driven startup of the klystron high-voltage power supply system.
- High Voltage Pad (HVPad) subsystem – Provides local control of the Variable Voltage Transformer (VVT) position, temperature, AC line, and cabinet monitoring.
- Klystrons subsystem – Manages klystron operating states, monitors inputs, and controls supporting subsystems, including cooling, ion pumps, magnets, and mod-anode circuits.
- Master Interlock subsystem – Aggregates critical signals from multiple subsystems and enforces interlock chains to ensure safe RF operation.
- Feeder subsystem – Monitors water-cooling systems for circulators, circulator loads, and dummy loads, providing operational permits to the Master Interlock PLC based on configuration mode.

* Work supported by the Office of Science, Office of Basic Energy Sciences, of the U.S. Department of Energy under Contract No. DE-AC02-05CH11231
† nusaqib@lbl.gov

- Fast Interlock subsystem – Serves as a communication gateway between the Master Interlock PLC and the FPGA-based fast interlock board, implementing exchange of RF power and arc detection parameters.

## UPGRADE PLAN AND STATUS

Figure 2 illustrates the upgrade system architecture, which consolidates related subsystems and reduces the number of PLC CPUs. The control system is restructured into four subsystems: HVPS, Klystron 1, Klystron 2, and the Master Interlock. The upgrade is being implemented in multiple phases to minimize operational disruptions, reduce technical risks, and ensure a smooth transition to a modern, sustainable platform. The four planned sequential phases are as follows:

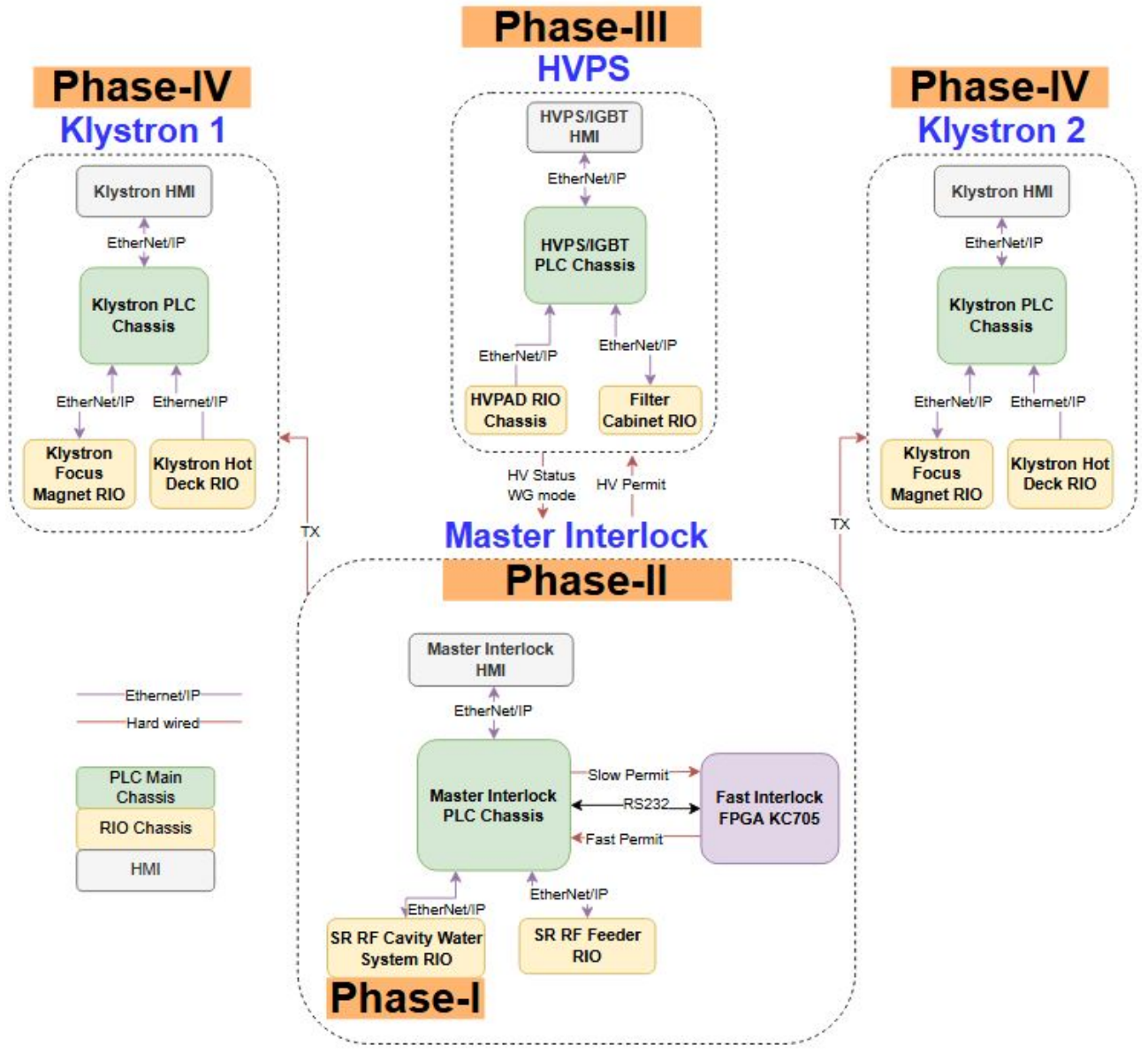

Figure 2: Upgrade system architecture.

### *Phase 1: Cavity Water Subsystem*

The first phase focused on replacing the legacy cavity water control system, transitioning from relay-based chassis hardware to a PLC-based solution. Details of this phase are presented in a separate conference paper [3]. This phase was successfully completed during the summer shutdown of 2024 and is currently operating smoothly.

### *Phase 2: Master Interlock Subsystem*

This phase replaces the Master Interlock subsystem. The Feeder and Fast Interlock subsystems are consolidated as remote I/O (RIO) chassis. This phase is currently in progress and is scheduled for completion during the summer shutdown 2025.

### *Phase 3: HVPS Subsystem*

The third phase upgrades the HVPS subsystem, while the closely associated HVPad subsystem is consolidated as a RIO chassis. This phase is scheduled for completion during the summer shutdown 2026.

### *Phase 4: Klystrons Subsystem*

The final phase replaces the Klystron subsystems and their associated RIO modules. This phase is scheduled for completion during the summer shutdown 2027.

The PLC and RIO chassis planned for the upgrade include:

- Phase I:
  - Cavity Water Control PLC chassis
- Phase II:
  - Master Interlock PLC chassis
  - Feeder RIO chassis
  - Cavity Water Control RIO chassis (Phase I PLC chassis with CPU replaced by a remote I/O adapter)
  - Fast Interlock RIO chassis
- Phase III:
  - HVPS PLC chassis
  - HVPad RIO chassis
  - Filter Cabinet RIO chassis
- Phase IV:
  - Klystron 1 PLC chassis
  - Klystron 1 Hot Deck RIO chassis
  - Klystron 1 Focus Magnet RIO chassis
  - Klystron 2 PLC chassis
  - Klystron 2 Hot Deck RIO chassis
  - Klystron 2 Focus Magnet RIO chassis

## SYSTEM DESIGN

The upgrade prioritizes maintainability, scalability, and consistency across the Storage Ring RF control system. The following subsections present an overview of the upgraded system's hardware and software design.

### *Hardware*

The Allen-Bradley CompactLogix 5380 series was selected as the upgrade platform to ensure compatibility with the PLC-based systems under development for ALS-U. Each PLC chassis is designed with front-mounted PLC modules and rear-panel connectors, supplemented by accessible terminal blocks on both sides to facilitate troubleshooting and future modifications. A standardized layout is applied across all chassis to improve maintainability and consistency. Figures 3 and 4 show the master interlock chassis: the front view highlights the modules and terminal blocks, while the rear view illustrates the subsystem-organized connectors and additional terminal blocks.

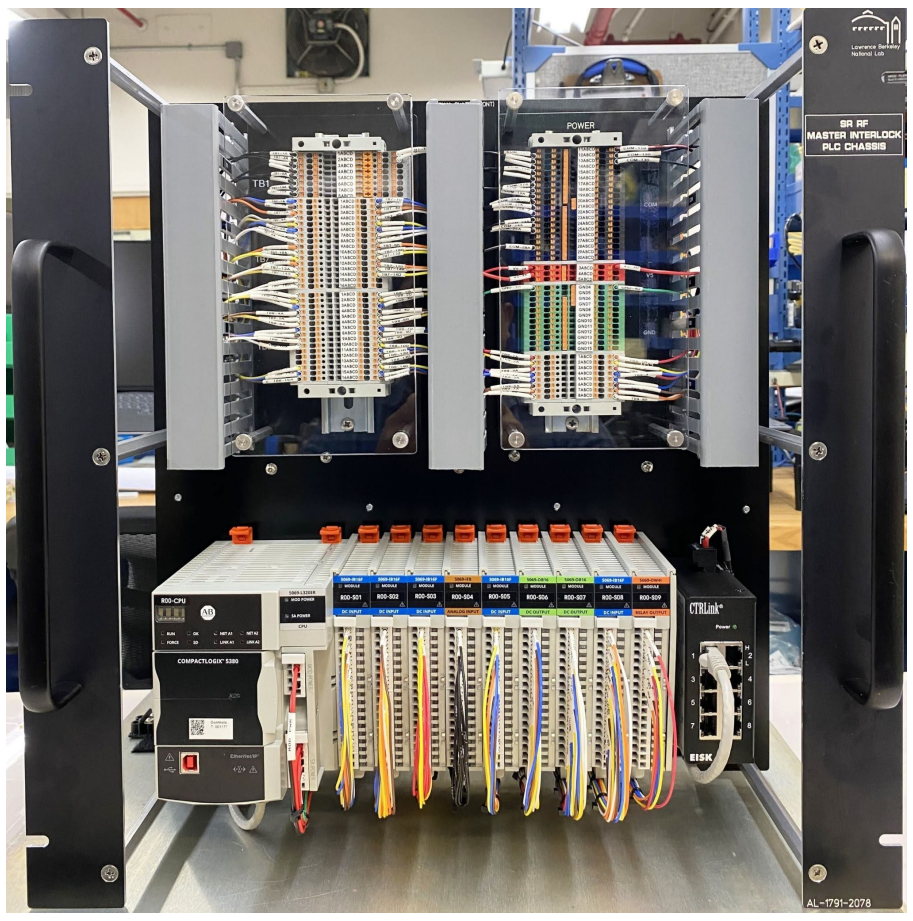

Figure 3: Front view of the master interlock chassis.

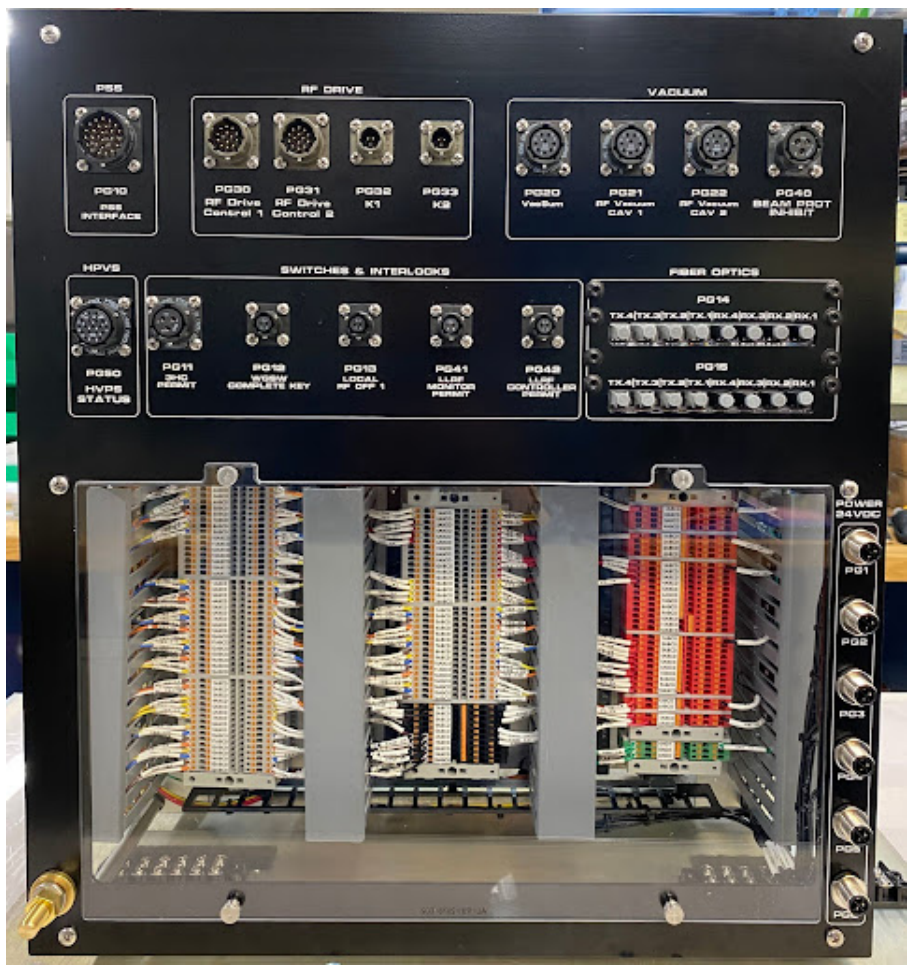

Figure 4: Rear view of the master interlock chassis.

## *Software*

The software consists of two layers: the PLC layer and the EPICS layer. For the PLC layer, Studio 5000 is used as the development environment. The PLC code is structured using Add-On Instructions (AOIs) and User-Defined Types (UDTs) to improve scalability, maintainability, and code consistency.The following subsections list some of the standard AOIs and base UDTs that are applicable across all PLC programs.

### Add-On Instructions

- Calibration AOI: Applies linear regression to analog signals for accurate scaling and calibration.
- Latch AOI: Latches a Boolean fault signal for persistent fault monitoring.
- Threshold AOI: Compares analog signal values with warning and trip thresholds to generate corresponding Boolean warning and trip status signals.
- First Fault AOI: Latches the first fault bit corresponding to the initial signal that caused a system fault.

### User-Defined Types

- UDT bi: Contains raw status, latched status, bypass, and first-fault bits for a Boolean signal.
- UDT ai: Contains raw and scaled values, warning and trip thresholds, gain and offset calibration factors, warning and trip Boolean status, bypass, and first-fault bits for an analog signal.
- UDT HMI: Contains commands received from the HMI interface.
- UDT EPICS: Contains commands received from EPICS.

The HMI screens follow the hierarchical structure of the PLC UDTs. Studio 5000 View Designer Add-On Graphics (AOGs) are created for commonly used UDTs, enabling modular and maintainable displays. Figure 5 shows an example screen integrating these AOGs to present the status, control, and configuration of system components.

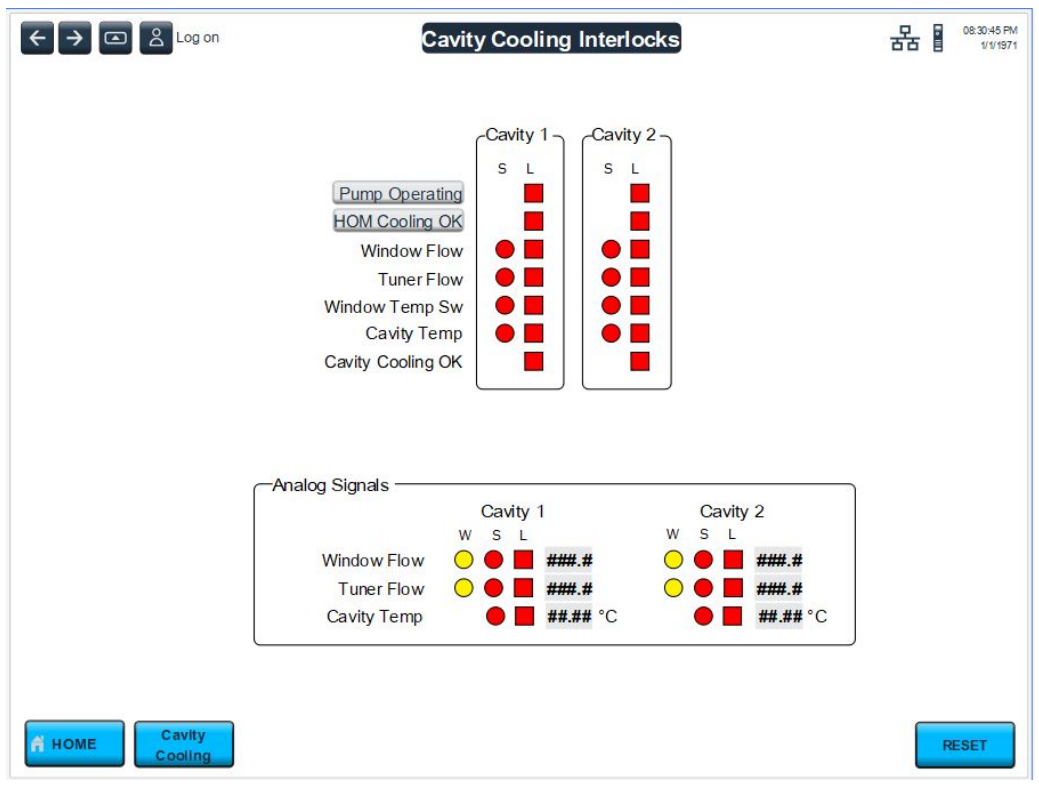

Figure 5: HMI screen.

EPICS IOCs interface with the PLCs using the Ethernet/IP protocol. The database is organized to reflect the hierarchical structure of the PLC UDTs, ensuring consistent mapping between PLC tags and EPICS records. Base UDTs are implemented as EPICS database templates, while all high-level UDTs and other controller tags are consolidated into a substitution file. A Python tool generates this substitution file by converting a Google Sheet containing the PVs and corresponding controller tags.

The Phoebus Operator Interfaces (OPIs) are developed using a hierarchical approach that mirrors the HMI screen structure, ensuring consistency between operator displays and local HMIs. Template OPIs, corresponding to HMI Add-On Graphics, are embedded into higher-level OPIs, enabling modular and reusable display components. Figure 6 illustrates an OPI constructed from these display templates, corresponding to the HMI screen described earlier.

# CONCLUSION

The Storage Ring RF control system upgrade at ALS represents a critical step toward ensuring long-term reliability and maintainability. Replacing the legacy, end-of-life Horner PLCs with Allen-Bradley CompactLogix PLCs provides a modern, scalable, and industry-standard platform capable of supporting future expansions and system enhancements.

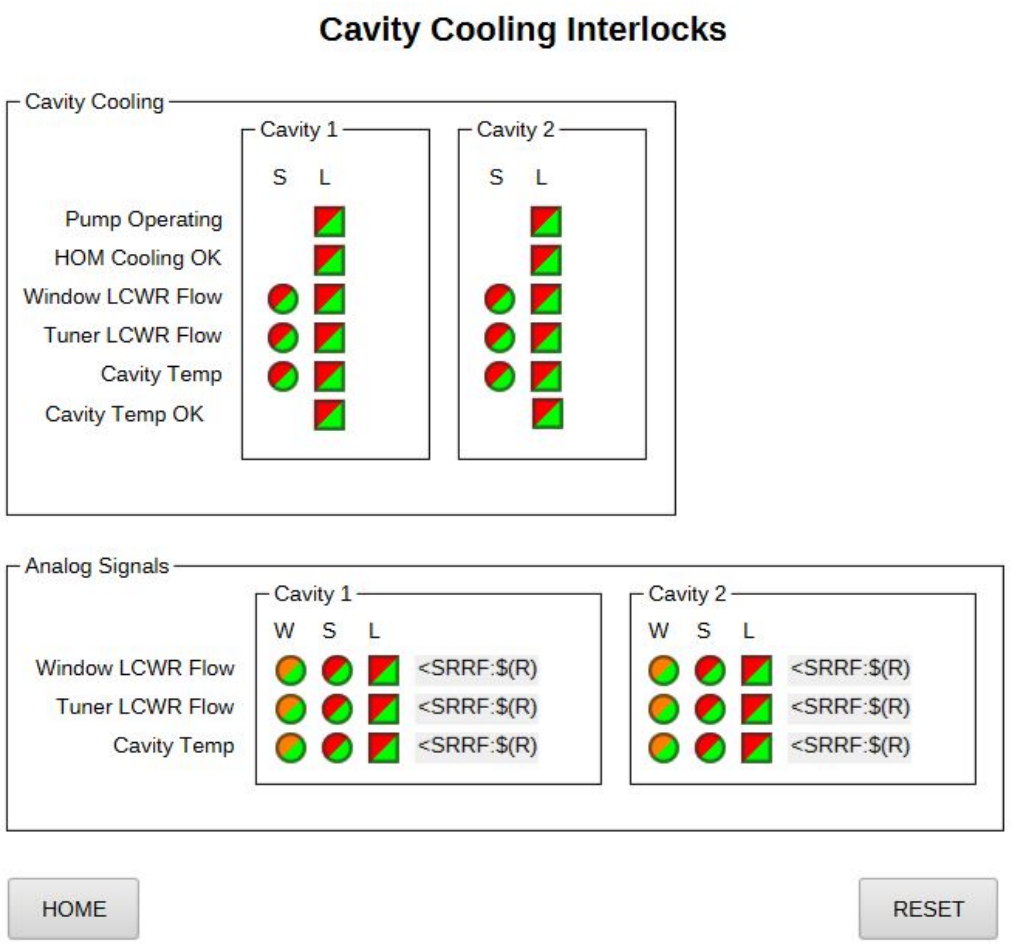

Figure 6: Phoebus OPI.

Phase I, which addressed the Cavity Water System, has been successfully completed and is operating smoothly. Work on the Master Interlock system is currently underway, with completion scheduled for the 2025 summer shutdown, followed by the HVPS and Klystron systems in subsequent summer shutdowns.